\documentclass[namedreferences]{solarphysics}
\usepackage[optionalrh]{spr-sola-addons}
\usepackage{threeparttable}
\usepackage{graphicx}
\usepackage{graphics}
\usepackage{color}   
\usepackage{url}    

\newcommand{\arcsec}{^{\prime\prime}}

\begin{document}
\begin{article}
\begin{opening}

\title{Temporal Pointing Variations of The Solar Dynamics Observatory's HMI and AIA Instruments on Sub-Weekly Time Scales}
%

%
\author{N. Brice Orange, Hakeem M. Oluseyi, David L. Chesny, Maulik Patel, Patrick Champey, Katie Hesterly, Dylan Anthony, and Robert Treen}
\runningtitle{AIA and HMI Co-registration}
\runningauthor{Orange {\it et al.}}
\institute{Department of Physics \& Space Sciences, Florida Institute of Technology, Melbourne, FL 32901, USA}

\begin{abstract}
Achieving sub-arcsecond co-registration across varying time-lines of multi-wavelength and instrument images is not trivial, and requires accurate characterization of instrument pointing jitter. In this work we have investigated internal pointing errors, on daily and yearly time-scales, occurring across the \textit{Solar Dynamics Observatory}'s (SDO) {\it Atmospheric Imaging Assembly} (AIA) and { \it Helioseismic Magnetic Imager} (HMI). Using cross-correlation techniques on AIA 1700\,{\AA} passband and HMI line-of-sight (LOS) magnetograms, from three years of observational image pairs at approximately three day intervals, internal pointing errors are quantified. Pointing variations of $\pm$\,0.26$\arcsec$ (jitter limited) and $\pm$\,0.50$\arcsec$ in the solar East-West ($x$) and North-South ($y$) directions, respectively, are measured. AIA observations of the Venus June 2012 transit are used to measure existing coalignment offsets in all passbands. We find AIA passband pointing variations are $\langle \Delta X_{CO} \rangle$\,$=$\, 1.10$\arcsec$\,$\pm$\,1.41$\arcsec$ and $\langle \Delta Y_{CO} \rangle$\,$=$\, 1.25$\arcsec$\,$\pm$\,1.24$\arcsec$, when aligned to HMI's nominal image center, referred to herein as the CutOut technique (CO). Minimal long-term pointing variations found between limb and correlation derived pointings provide evidence that image center positions provided by the instrument teams achieve single pixel accuracy on time-scales below their characterization. However, daily AIA passband pointing variations of $\lesssim$\,1.18$\arcsec$ indicate autonomous sub-arcsecond co-registration is not yet fully achievable.
\end{abstract}

\end{opening}

\section{Introduction}
\label{sec:intro}
Addressing the coronal heating problem is not trivial, as it requires the use of high resolution multi-wavelength and temporal images with sub-arcsecond co-registration. Currently, in the solar physics field the {\it Solar Dynamics Observatory}'s (SDO) {\it Atmospheric Imaging Assembly} (AIA; \opencite{Lemenetal2012}) and {\it Heliographic Magnetic Imager} (HMI; \opencite{Schouetal2012SoPh}) instruments are providing unprecedented amounts, approximately one image a second, of high resolution data ($\lesssim$\,0.6$\arcsec$). AIA takes full-disk images of the Sun in the following nine passbands: 94\,{\AA} ($\log T$\,$\approx$\,6.8), 131\,{\AA} ($\log T$\,$\approx$\,5.8), 171\,{\AA} ($\log T$\,$\approx$\,5.9), 193\,{\AA} ($\log T$\,$\approx$\,6.2, 7.2), 211\,{\AA} ($\log T$\,$\approx$\,6.3), 304\,{\AA} ($\log T$\,$\approx$\,4.8), 335\,{\AA} ($\log T$\,$\approx$\,6.4), 1600\,{\AA} ($\log T$\,$\approx$ 5.0), 1700\,{\AA} ($\log T$\,$\approx$\,3.7), and 4500\,{\AA} ($\log T$\,$\approx$\,3.7; observed typically every $\approx$\,30 min), while HMI takes full-disk images of the Sun's line-of-sight (LOS) magnetic field ($\approx$\,45 s).

Some recent studies have highlighted that commonly used multi-passband alignments for AIA and HMI, considered to yield sub-pixel ($\le$\,0.6$\arcsec$) accuracy, deliver a less than optimized status ($\approx$\,1$\arcsec$; \opencite{DelZannaetal2011}; \opencite{Brooksetal2012ApJ}). These alignment techniques, both of which utilize {\sf aia\_prep.pro}, {\it i.e.}, standard Solar SoftWare (SSW), and result in level-1.5 data, are summarized as follows: aligning multi-passband charge-couple device (CCD) centers as Sun center (SC) or using a single passband's documented alignment information as a fiducial reference position for all other passbands. Thermal jitter motion, which places a systematic limit on achievable co-alignment accuracy, reported to affect each AIA telescope is $\approx$\,0.3$\arcsec$ \cite{Lemenetal2012,Aschwandenetal2011}, while that of HMI remains un-reported (as of time of writing; \opencite{Schouetal2012SoPh}).

\citeauthor{Shineetal2011SPD} (\citeyear{Shineetal2011SPD}) showed AIA passband image pointings are not temporally stable, suggested as a result of thermal flexing, for the four AIA telescopes and indicated monitoring and characterization of each telescope's nominal pointing ({\it i.e.}, offset relative to solar center) is required. Currently, {\it per} AIA passband, image pointings are defined from an automated limb finder routine that performs an interactive Hough transform on candidate limb pixels to converge on the solar center ($x$, $y$) and define it's radial distance \cite{Shineetal2011SPD}. A daily average of these image center positions, performed once a week, are used to define the master pointings that are propagated in documented header information \cite{Shineetal2011SPD}. However, image positions change significantly on weekly time scales, and daily variations ($\gtrsim$\,0.6$\arcsec$) are not characterized by master pointings \cite{Shineetal2011SPD}. AIA image plate scales and angles are considered significantly more stable than that of the aforementioned offsets, with \citeauthor{Shineetal2011SPD} (\citeyear{Shineetal2011SPD}) indicating no evidence of scale change and angle variations of 0.001$^\circ$. HMI's master pointings are defined from similar techniques and time scales as those used by AIA.

\begin{figure}[!t]
\begin{center}
 \includegraphics[scale=0.6]{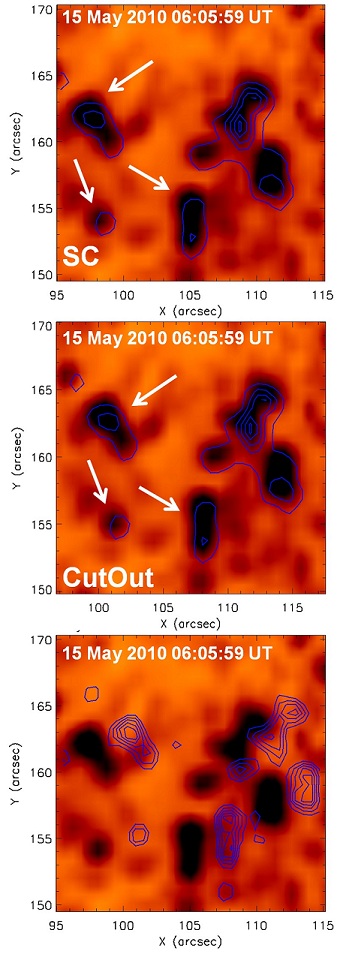}
 \caption{The top and middle panels (pre-processed under SC and CutOut techniques, respectively) are AIA 1700\,{\AA} intensity images (DN pix$^{-1}$ s$^{-1}$), shown on inverted color scale, with contours showing regions of positive magnetic flux (blue solid lines) derived from HMI's LOS magnetograms and plotted at levels of $\pm$\,25\,--\,100 G. Bottom panel is AIA 1700\,{\AA} same as that in the top but contours are from the middle panel image, which shows the difference between the pre-processing techniques is a solar center offset ($\approx$\,5$\arcsec$ in both the $x$ and $y$ directions).}
\label{fig:AIA1700HMI_201005_CCDCUTOUT_Zoom}
\end{center}
\end{figure}

Using cross-correlation techniques on AIA and HMI observational image pairs, performed on sub-weekly time-scales, we measure instrument jitter in the master pointing data by directly comparing limb and correlated image center positions. The June 2012 Venus transit is used to measure pointing variations of each of AIA's passbands. Throughout the remainder of this paper the term SC technique refers to multi-passband image CCD centers defined as Sun center, while CutOut is that where HMI's master pointing is as the fiducial reference position for all AIA passbands. It is noted, reported co-registration discrepancies are consistent across both co-alignment techniques summarized above, but result in a constant solar center offset $\approx$\,5$\arcsec$ (Figure~\ref{fig:AIA1700HMI_201005_CCDCUTOUT_Zoom}) between the two (Boerner 2013, private communication).

In that respect, the remainder of this paper is outlined as follows: in Section~\ref{sec:quantify_aiahmi_errs} we provide a long-term, sub-weekly, characterization of AIA and HMI instrument jitter and compare image pointings resultant from limb fitting and cross-correlation techniques; Section~\ref{sec:aia_internal_investigation} quantifies AIA passband pointing variations {\it via} observational data from the June 2012 Venus transit; a discussion of our findings is provided in Section~\ref{sec:Discussion}; and concluding remarks are given in Section~\ref{sec:concl_summary}.

\section{AIA and HMI Sub-Weekly Pointing Errors}\label{sec:quantify_aiahmi_errs}

Measurements of AIA and HMI's instrument pointing variations are performed at approximately three day intervals spanning May 2010\,--\,March 2013 from cross-correlated image pairs of the 1700\,{\AA} passband (which directly observes the solar photosphere) and LOS magnetograms. Cross-correlation, widely considered to yield sufficient accuracy ($\lesssim$ 1 pixel; {\it e.g.}, \citeauthor{Brooksetal2012ApJ} \citeyear{Brooksetal2012ApJ}, \citeauthor{DelZannaetal2011} \citeyear{DelZannaetal2011}) is done using visually bright point-like features or magnetic neutral lines (Figure~\ref{fig:AIA1700HMI_201005_CCDCUTOUT_Zoom}).

Image pairs were pre-processed to level-1.5 using {\sf aia\_prep.pro} (which included roll angle corrections), CutOut aligned, and magnetograms interpolated to the effective resolution of FUV images ($\approx$\,0.6$\arcsec$). Solar rotation effects, $\approx$\,0.17 arcsec min$^{-1}$ \cite{SarroBerihuete2011} close to disk center, were minimized by using observational time differences below AIA's thermal jitter motion ({\it i.e.}, $\le$\,2 min), and  the standard SSW routine {\sf drot\_map.pro}. FUV and magnetogram images were then mapped to solar coordinates by calibrating each image's axes with respect to actual Stonyhurst heliographic coordinates, {\it i.e}, coordinate (0,0) corresponds to solar center. Disk regions of $\pm$\,300$\arcsec$ from solar center are only considered to minimize projection effects. {\it Per} observational image pair, two measurements of the $x$ and $y$ co-alignment offsets, relative to HMI's master pointings ($R^{{\rm obs}}_x$ and $R^{{\rm obs}}_y$, respectively), were made on regions comprised of bright network and small compact features ({\it e.g.}, similar to those observed in Figure~\ref{fig:AIA1700HMI_201005_CCDCUTOUT_Zoom}). The resultant $R^{{\rm obs}}_x$ and $R^{{\rm obs}}_y$ offsets were then smoothed {\it per} day with errors propagated from the differences in daily measurements.
\begin{figure}[!t]
\begin{center}
 \includegraphics[scale=0.22]{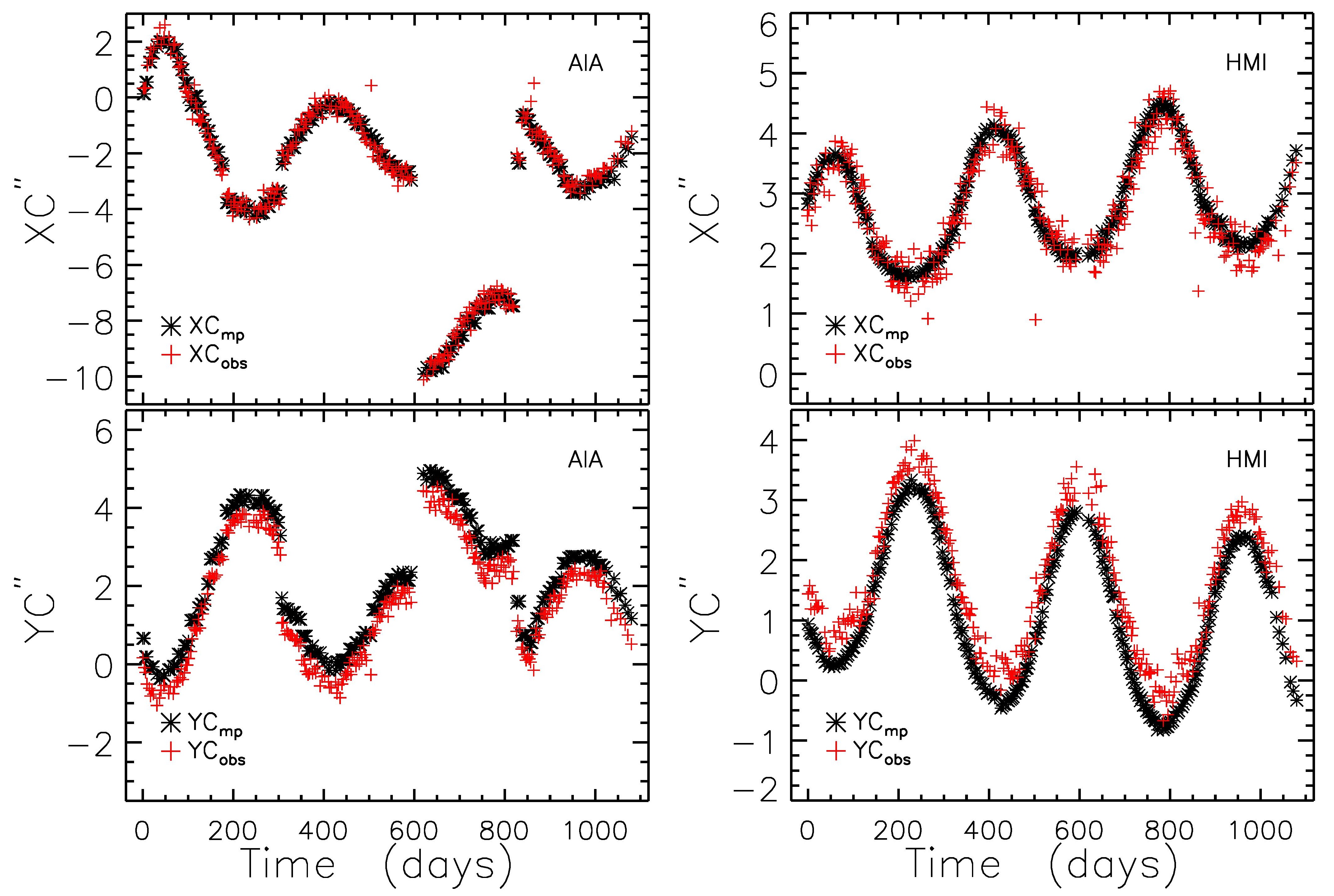}
 \caption{{\rm Top}, from left to right, are plots of AIA and HMI master pointings, respectively, ($XC_{{\rm mp}}$; asterisks) provided by the SDO teams and observed image centers from cross-correlation techniques ( $XC_{{\rm obs}}$; red pluses) in the solar $x$ direction {\it versus} time (May 2010\,--\,March 2013), respectively. {\rm Bottom}, similar to top panels for pointings in the solar $y$ direction.}
\label{fig:AIA1700_HMI_XYMPS_RES_VSTIME}
\end{center}
\end{figure}

\begin{table}[!t]
\caption{Observational date range, image time difference ($\langle$\,$\Delta$\,${t}$\,$\rangle$; in seconds), typical observed $x$ and $y$ image center offsets ($\langle$\,$R^{{\rm obs}}_x$\,$\rangle$ and $\langle$$R^{{\rm obs}}_y$\,$\rangle$, respectively, in arcsec) for co-aligned 1700\,\AA~far-ultraviolet (FUV) images and LOS magnetograms.}
\label{tbl:AIAHMI_COREG_STATS}
\vspace{0.5mm}
\begin{tabular}{cccc}
\hline
Obs. Date Range & $\langle$\,$\Delta$\,${t}$\,$\rangle$ & $\langle R^{{\rm obs}}_x \rangle$ & $\langle R^{{\rm obs}}_y \rangle$ \\
\hline
May 2010 -- March 2013 & 5.71 & 0.26 & 0.50 \\
\hline
\end{tabular}
\end{table}

Figure~\ref{fig:AIA1700_HMI_XYMPS_RES_VSTIME} displays AIA and HMI's master pointings ({\it i.e.,} $XC/YC_{{\rm mp}}$ derived from limb fitting techniques) compared to the pointings obtained from cross-correlating image pairs ({\it i.e.}, $XC/YC_{{\rm obs}}$) at sub-weekly time-scales. A summary of our co-alignment offsets are provided in Table~\ref{tbl:AIAHMI_COREG_STATS}, with typical values of $\approx$\,0.3$\arcsec$ and 0.5$\arcsec$ (in solar $x$ and $y$ directions, respectively). These results are consistent with reported mis-alignments ($\lesssim$ 1$\arcsec$; \opencite{DelZannaetal2011}; \opencite{Brooksetal2012ApJ}) as well as those expected from daily variations ($\lesssim$ 1.2$\arcsec$; \opencite{Shineetal2011SPD}). Moreover, $R^{{\rm obs}}_x$ is within the limits of instrumental jitter, and as observed in Figure~\ref{fig:AIA1700_HMI_XYMPS_RES_VSTIME} is indicative of random scatter. Solar $y$ pointing offsets varied minimally ($\approx$\,$\pm$\,0.02$\arcsec$) over the three year span, and indicate their alignment agreement could be autonomously improved (Figure~\ref{fig:AIA1700_HMI_XYMPS_RES_VSTIME}).

\section{Using the Venus Transit for Investigation of Internal AIA Co-Registration Errors}\label{sec:aia_internal_investigation}

\begin{figure}[!t]
\begin{center}
 \includegraphics[scale=0.6]{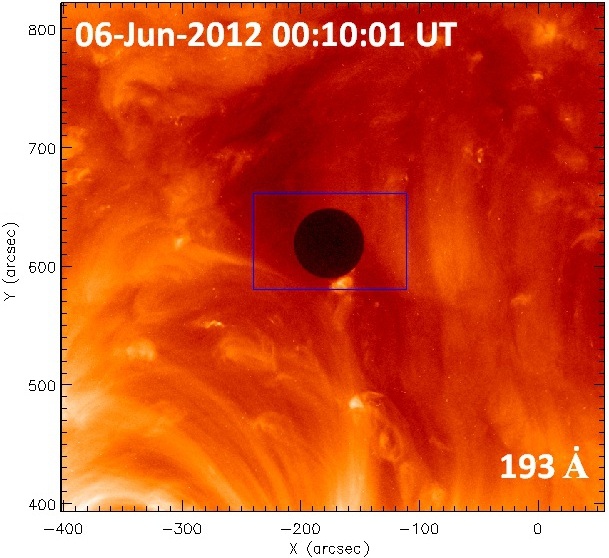}
 \caption{AIA 193\,{\AA} passband observed on 6 June 2012 at 00:10:01 UT. The image is displayed on a $\log_{10}$ scale and the boxed blue region surrounds the disk of Venus. }
\label{fig:VT_OffOn_Images}
\end{center}
\end{figure}

Investigation of daily AIA pointing variations uses observational data of the 5-6 June 2012 Venus transit (23:00:00 UT\,--\,04:00:00 UT), while Venus' silhouette was fully visible on the solar disk (Figure~\ref{fig:VT_OffOn_Images}). Venus' transit provides the ideal fiducial reference for measuring such errors since its silhouette must be seen at identical locations in each passband when the transit speed across the solar surface is accounted for \cite{Kamioetal2011,Shimizuetal2007}.

\begin{figure}[!t]
\begin{center}
 \includegraphics[scale=0.37,angle=90]{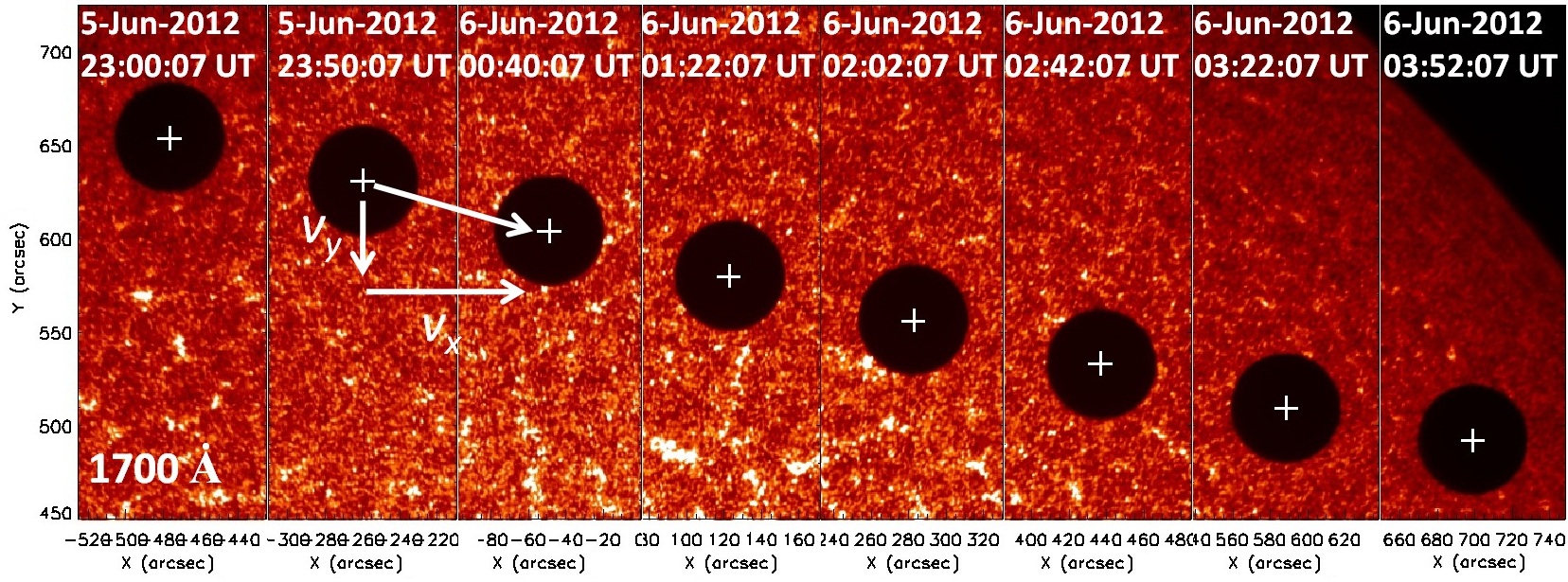}
 \caption{Sequence of AIA 1700\,{\AA} images, observed on 5-6 June 2012 from 23:00:07 UT -- 03:52:07 UT (left to right, respectively) with pluses denoting Venus' centroid, defined by the technique discussed in the text (Section~\ref{sec:VT_analysis}). Arrows are shown to represent the running difference technique applied to the data to measure the planet's transit velocity in the $x$ and $y$ directions and provide fiducial references as a function of time from which co-alignment discrepancies between all other AIA passbands are measured.}
\label{fig:VT_RunDiff_Image}
\end{center}
\end{figure}

Daily pointing variations from the various AIA passbands are determined from a direct comparison of the heliographic coordinates of Venus' center to those predicted by deriving the Venus' velocity using running difference images (1700\,{\AA} passband; see Figure~\ref{fig:VT_RunDiff_Image}). The details of our observational data and method as well as results are described in the next two Sections~3.1 and 3.2.

\subsection{Venus Transit Analysis}\label{sec:VT_analysis}

The Venus transit data were pre-processed to obtain both SC and Cut-Out multi-image alignments, as well as those aligned to their respective master pointings (MP). For each pre-processing technique, a ten minute temporal cadence was used which resulted in $\approx$\,30 images of each passband with exception of 4500\,{\AA} images (5 observations). Custom written software, discussed in depth below, was then used to measure the heliographic coordinates of Venus' disk centroid from all images.

\begin{figure}[!t]
\begin{center}
 \includegraphics[scale=0.17]{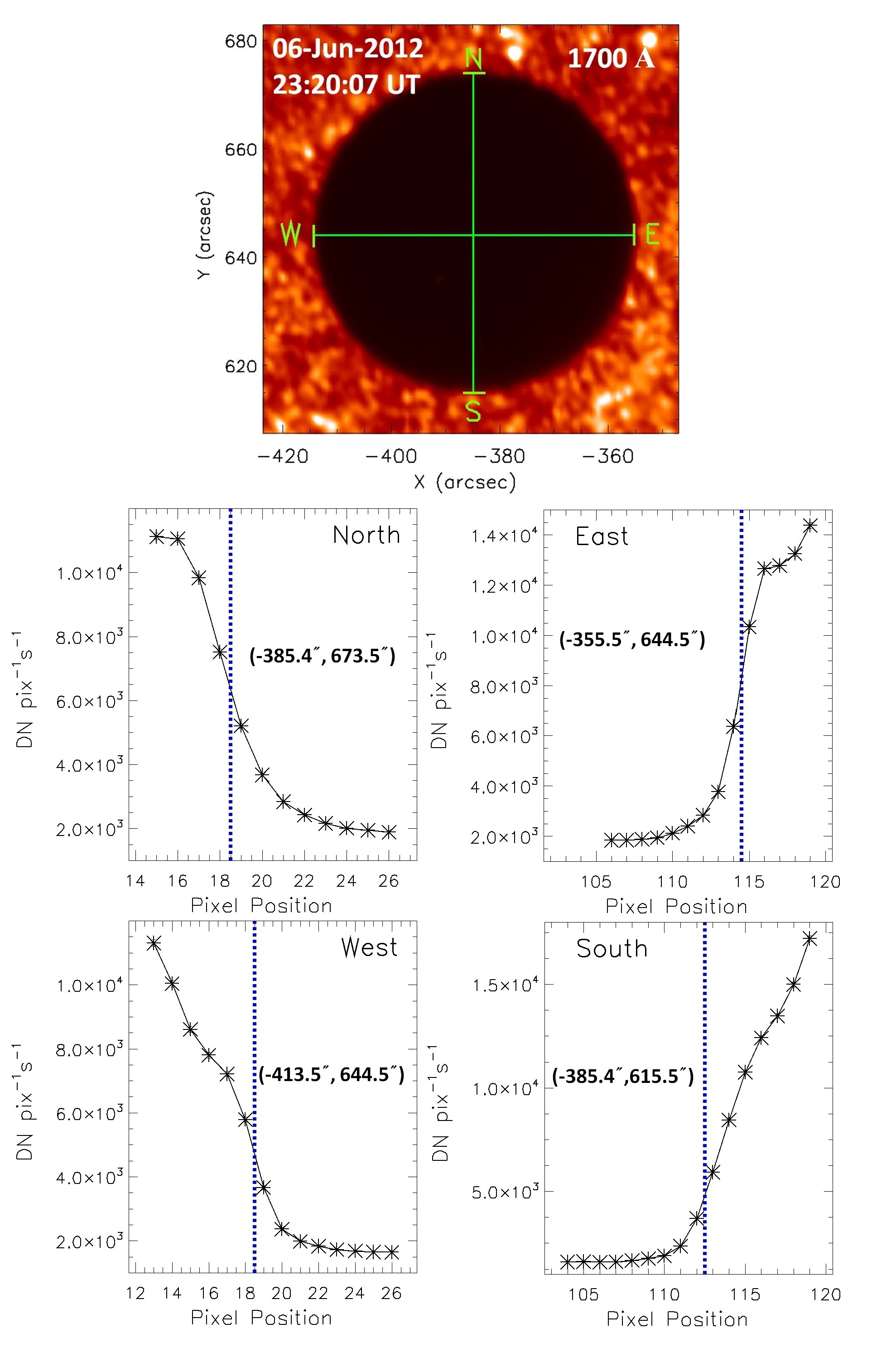}
 \caption{The {\rm top} panel is an intensity image centered on the silhouette of Venus's disk from AIA's 1700\,{\AA} passband observed on 5 June 2012 23:20:07 UT, with the center of the crosshairs representing the measured center of Venus' disk. The four {\rm bottom} panels are flux {\it versus} pixel position for the North, East, South, and West positions noted in the top panel from top left to bottom left in clockwise direction, respectively. Note, vertical dotted blue lines denote measured limb positions, as outlined in the text, of (-385.4$\arcsec$, 673.5$\arcsec$), (-385.4$\arcsec$, 673.5$\arcsec$), (-355.5$\arcsec$, 644.5$\arcsec$), and (-413.5$\arcsec$, 644.5$\arcsec$) for the North, East, South, and West positions, respectively, presented here in the form of solar $x$ and $y$ directions, respectively.}
\label{fig:VT_CENT_Tech}
\end{center}
\end{figure}

The software measured individually the vertical, North -- South (N--S), and horizontal, East -- West (E--W), disk centroid positions {\it vi} a calculating and comparing observed disk diameters to the planet's known angular size (58.3$\arcsec$; \opencite{Odenwald2012}). Calculation of the disk diameter required user selection of {\it loci} regions corresponding to a single orientation's limb position ({\it e.g.}, leading and trailing edges for E--W, see Figure~\ref{fig:VT_CENT_Tech}). {\it Per loci} region, limb positions along the vertical or horizontal for E--W and N--S diameters, respectively, were determined by measuring where the steepest flux gradients occurred (Figure~\ref{fig:VT_CENT_Tech}). The position of steepest flux gradient occurring furthest away from disk center was stored and visually checked. Uncertainties in each of the aforementioned measurements were derived from the plate scale of a single pixel {\it per} passband (defined in the header of the FITS data file) and residuals from comparing measured angular sizes to the known value.
\begin{figure}[!t]
\begin{center}
 \includegraphics[scale=0.28]{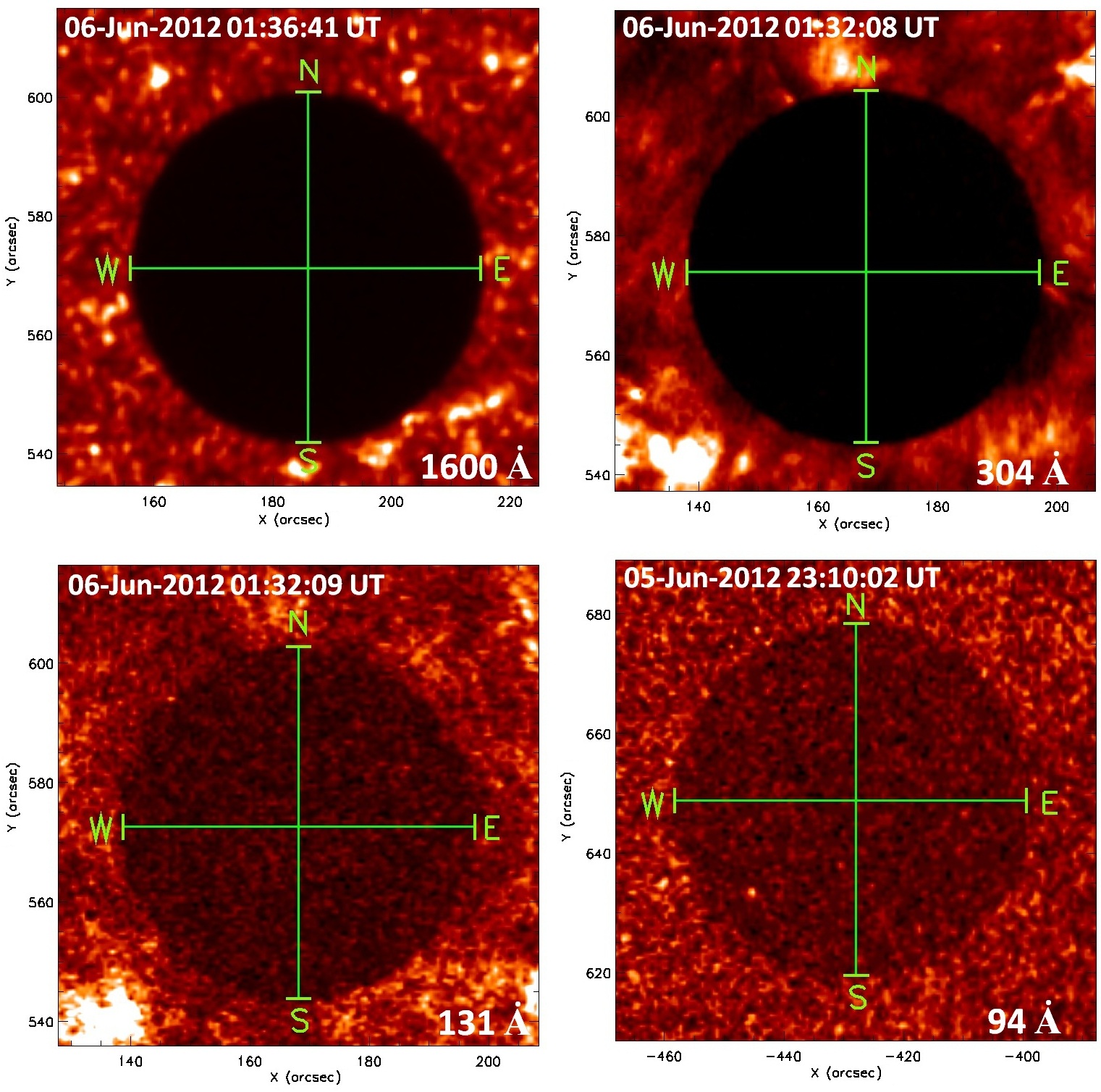}
 \caption{Intensity images centered on the silhouette of Venus' disk from AIA's 1600\,{\AA}, 304\,{\AA}, 94\,{\AA}, and 131\,{\AA} passbands from top right to bottom right in clockwise direction, respectively, observed on 5-6 June 2012 showing variations in the contrast of Venus' disk as observed by different passbands. On each image the center of the crosshairs represents Venus' measured centers determined from the techniques discussed in the text, respectively.}
\label{fig:VT_CENT_VaryingPassbands}
\end{center}
\end{figure}

Figure~\ref{fig:VT_CENT_Tech} provides an example of this entire process. Its left panel displays the resultant diameters (solid lines, plotted with a width of $\approx$\,1$\arcsec$ to be easily observable) and limb positions (marked by their respective direction and a solid line). The right panel displays flux {\it versus} position for the N--S and E--W disk edges corresponding to the aforementioned marked locations on the intensity image of the right panel. Figure~\ref{fig:VT_CENT_VaryingPassbands} has also been provided to illustrate the variations in Venus' contrast against the solar disk for a sample of passbands. It is noted, features which show up on Venus' disk in 131\,{\AA} and 94\,{\AA} images of Figure~\ref{fig:VT_CENT_VaryingPassbands} are directly indicative of the stray light which affects all AIA EUV passbands \cite{Poduvaletal2013ApJ}.

Using running differences of Venus' disk center position, its transit velocity relative to the solar surface was measured to be $v_x$\,$=$\,6.7$\times$10$^{-2}$\,$\pm$\,6.3$\times$10$^{-4}$ arcsec s$^{-1}$ and $v_y$\,$=$\,$-$9.2$\times$10$^{-3}$\,$\pm$\,6.3$\times$10$^{-4}$ arcsec s$^{-1}$ in the horizontal and vertical directions, respectively (Figure~\ref{fig:VT_RunDiff_Image}). Observational disk centers and the transit speed were then used to generate a set of predicted disk positions at one second intervals over the observational time frame used. Co-alignment offsets ($\Delta$\,$X$, $\Delta$\,$Y$), relative to 1700\,{\AA} images as function of transit time, were measured for co-aligned multi-passband data.

\subsection{Results}

\begin{figure}[!t]
\begin{center}
 \includegraphics[scale=0.23,angle=90]{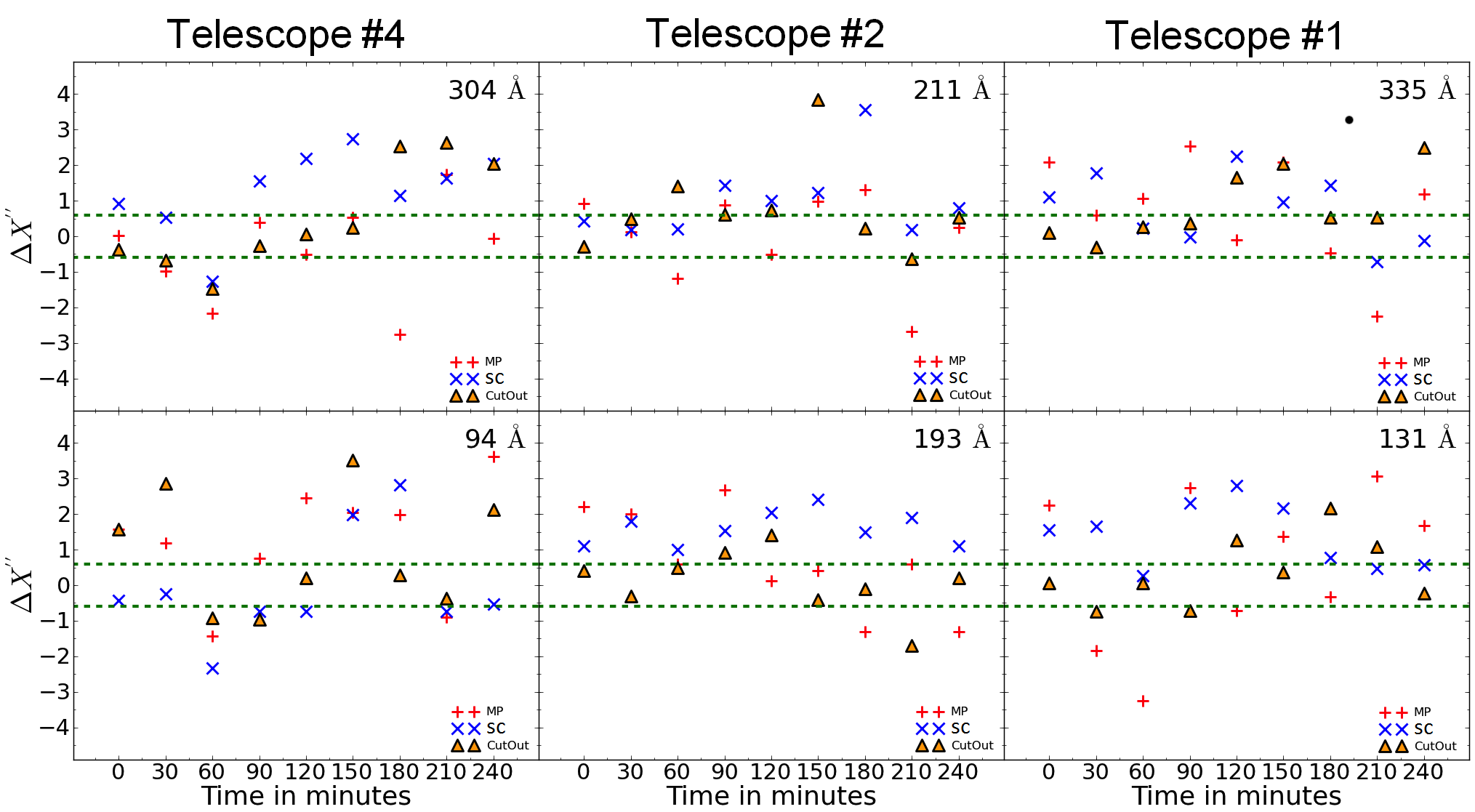}
 \caption{Solar $x$ co-alignment  offsets ($\Delta X$), relative to 1700\,{\AA} image positions, {\it versus} time (minutes measured relative to 5 June 2012 23:00:00 UT) shown as a function of AIA telescope number (1, 2, and 4 only; Lemen {\it et al.}, 2012) and smoothed over $\approx$\,30 min for visual clarity. Pre-processing techniques of MP, SC and CutOut are represented by red pluses, blue x-signs, and gold triangles on each plot, respectively.}
\label{fig:VT_XRESD_TELE_124}
\end{center}
\end{figure}

\begin{figure}[!t]
\begin{center}
 \includegraphics[scale=0.23,angle=90]{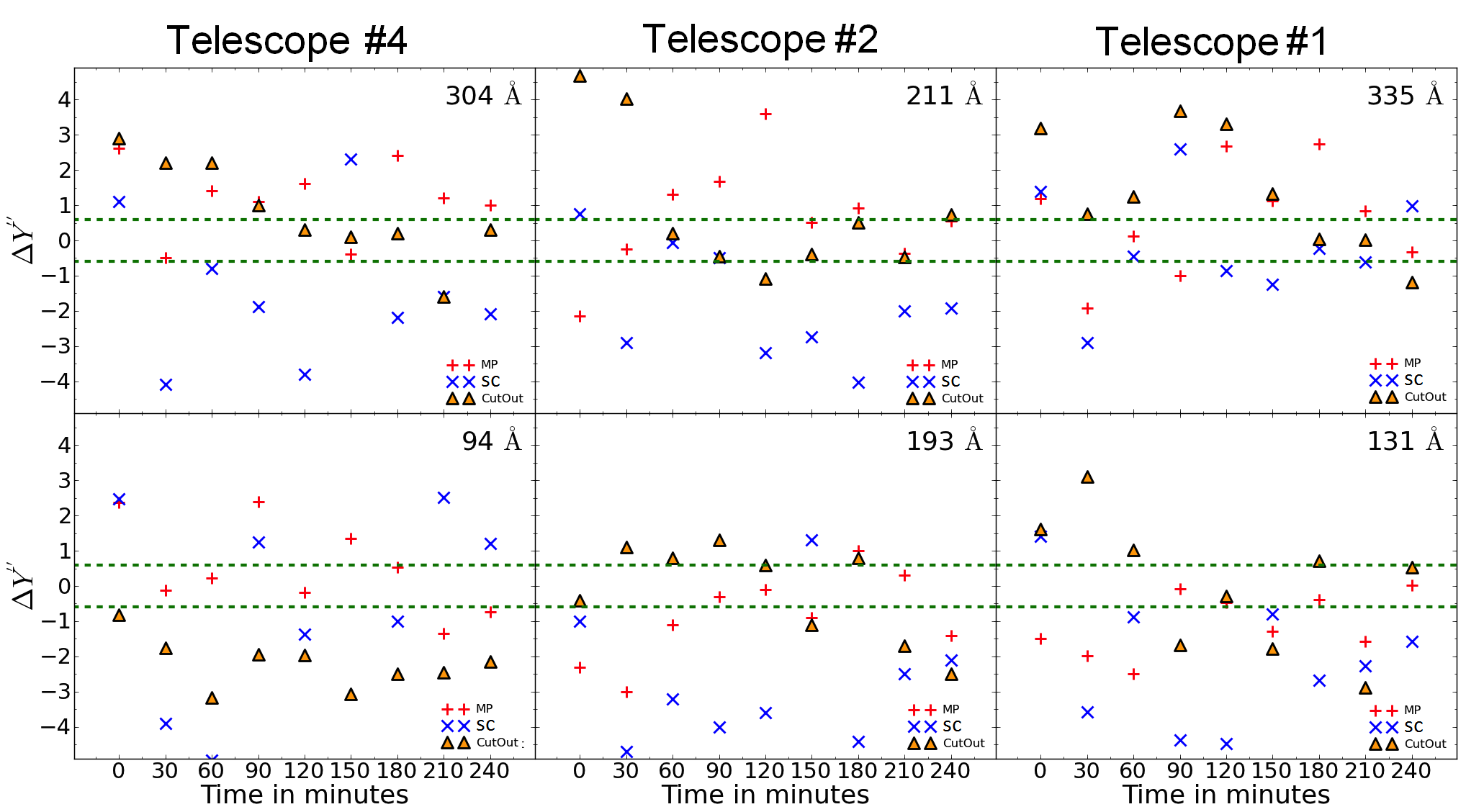}
 \caption{Solar $y$ co-alignment offsets ($\Delta Y$), relative to 1700\,{\AA} image positions, {\it versus} time (minutes measured relative to 5 June 2012 23:00:00 UT) shown as a function of AIA telescope number (1, 2, and 4 only; Lemen {\it et al.}, 2012) and smoothed over $\approx$\,30 min for visual clarity. Pre-processing techniques of MP, SC and CutOut are represented by red pluses, blue x-signs, and gold triangles on each plot, respectively.}
\label{fig:VT_YRESD_TELE_124}
\end{center}
\end{figure}

\begin{figure}[!t]
\begin{center}
 \includegraphics[scale=0.28]{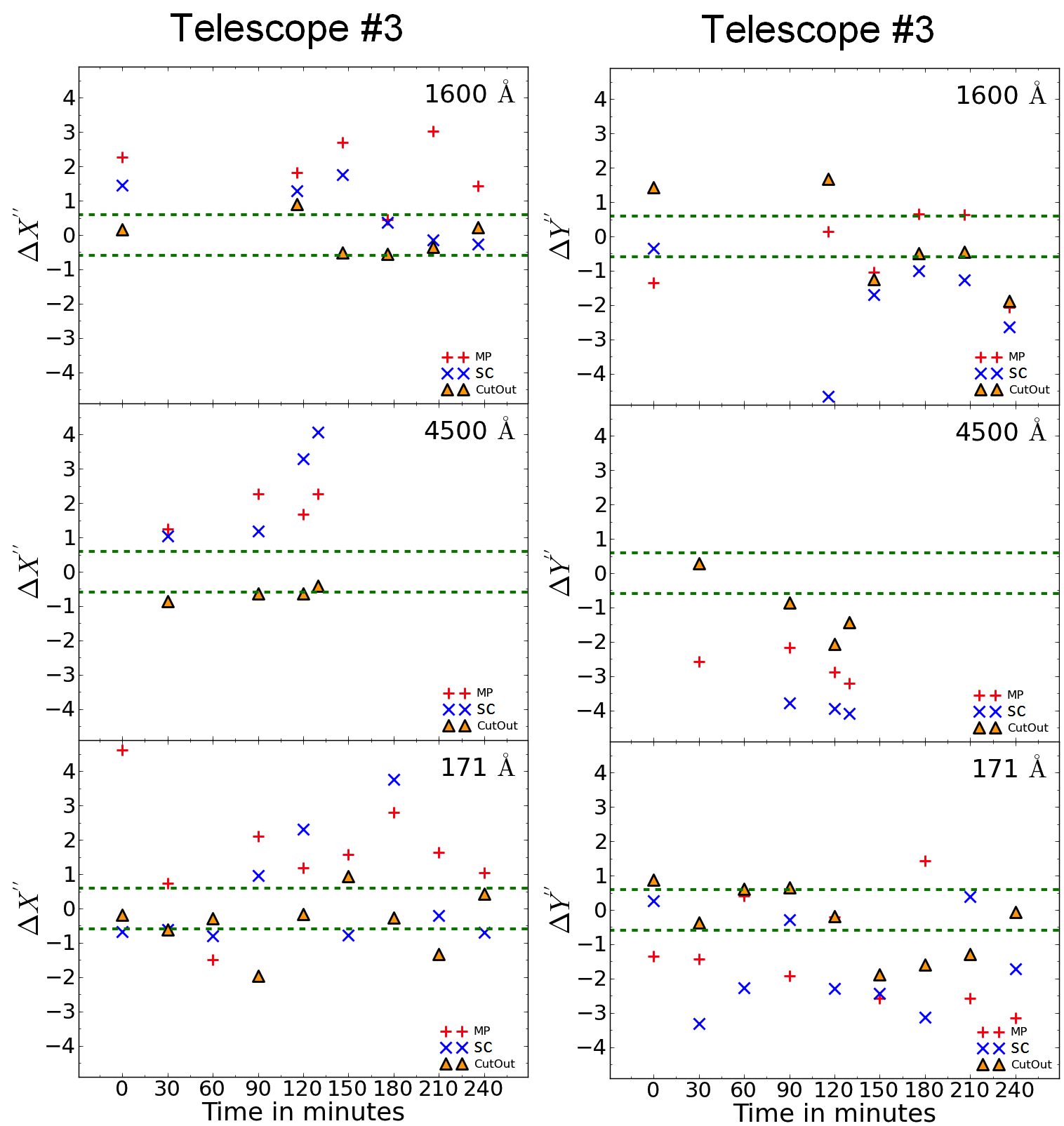}
 \caption{Solar $x$ and $y$ co-alignment offsets ($\Delta X$ and $\Delta Y$ in left and right columns, respectively), relative to 1700\,{\AA} image positions, {\it versus} time (minutes measured relative to 5 June 2012 23:00:00 UT) shown for AIA telescope number 3 (Lemen {\it et al.}, 2012) and smoothed over $\approx$\,30 min for visual clarity. Pre-processing techniques of MP, SC and CutOut are represented by red pluses, blue x-signs, and gold triangles on each plot, respectively.}
\label{fig:VT_XYRESD_TELE_3}
\end{center}
\end{figure}

Figures~\ref{fig:VT_XRESD_TELE_124} and \ref{fig:VT_YRESD_TELE_124} provide the $x$ and $y$ co-alignment offsets, respectively, (plotted at $\approx$\,30 min intervals for clarity) of AIA passbands imaged by telescopes 1, 2, and 4 \cite{Lemenetal2012}. Figure~\ref{fig:VT_XYRESD_TELE_3} displays both the $x$ and $y$ co-alignment offsets of AIA passbands imaged by telescope 3 \cite{Lemenetal2012}. As observed in these figures, significant variance exists, not only between disk center positions but as a function of time, too. Most typically, the $x$ and $y$ pointing variations are indicative of random scatter, but a number of passbands do exhibit a quasi-periodic nature in the $x$ direction. Using the 304\,{\AA} passband the period of this variation was estimated to be $\approx$\,5 h. However, we note that it is only a crude approximation since variations are found in other passbands ($\approx$\,$\pm$\,30 min, Figures~\ref{fig:VT_XRESD_TELE_124} and \ref{fig:VT_XYRESD_TELE_3}) while our observation sequence did not cover a full period.


Table~\ref{tbl:AIA_PREPROCESS_ALIGNRESULTS} presents the co-alignment offsets, averaged over seven passbands, and uncertainties, defined as the standard error on the mean (SEM),  for each pre-processing technique. It is noted, these results are not weighted by measurements derived from 94\,{\AA} and 4500\,{\AA} images for the following reasons: the low count-rates of the 94\,{\AA} passband \cite{Boerneretal2012SoPh}, and the significantly longer temporal cadence of the 4500\,{\AA} passband.
\begin{table}[!t]
\caption{Typical co-alignment offsets, in arcsec, (averaged over seven AIA passbands, and measured relative to the 1700\,{\AA} image positions) and their resultant uncertainties, defined as the standard error on the mean (SEM), for pre-processing images under SC, MP, and CutOut assumptions.}
\label{tbl:AIA_PREPROCESS_ALIGNRESULTS}
\vspace{0.5mm}
\begin{tabular}{ccccc}
\hline
Pre-Process & $\langle X  \rangle$ & $\sigma ^{X} _{SEM}$	& $\langle Y \rangle$ & $\sigma ^{Y} _{SEM}$ \\
\hline
SC    & 1.47 & 2.86 &	1.83 & 2.29 \\
MP      & 1.29 & 1.60 &	1.37 & 1.66 \\
Cut-Out & 0.87 & 1.14 &	1.18 & 1.46 \\
\hline
\end{tabular}
\end{table}
The results of Table~\ref{tbl:AIA_PREPROCESS_ALIGNRESULTS} indicate that image pointings vary by $\gtrsim$\,2 pixel, independent of solar direction, on daily time-scales. We point out, no technique results in sub-pixel ($\lesssim$\,0.6$\arcsec$) alignment, as expected \cite{Shineetal2011SPD}, while resultant uncertainties indicate minimal variations of $\approx$\,1.3$\arcsec$ occur over a few hours.

Co-alignment offsets {\it per} passband were re-measured by letting individual passbands serve as the fiducial reference ({\it i.e.}, using running image differences to derive and predict the planet's disk center positions), again barring 94\,{\AA} and 4500\,{\AA} images. The results of this analysis are shown in Table~\ref{tbl:AIA_PREPROCESS_ALIGNRESULTS_ALLBANDS}. As observed, the typical alignment accuracy, as well as its respective uncertainty, is consistent with our previous reports for the 1700\,{\AA} passband. Therefore, using
our resultant co-alignment offsets, the typical AIA passband alignment accuracy, derived from eight passbands, is
\begin{equation}
\begin{array}{ll}
\langle \Delta X_{CO} \rangle = & 1.^{\prime \prime}10 \pm 1.^{\prime \prime}41 \hspace{0.1in} {\rm and,}\\
\langle \Delta Y_{CO} \rangle = & 1.^{\prime \prime}25 \pm 1.^{\prime \prime}24, \\
\end{array}
\end{equation}
in the $x$ and $y$ directions, respectively, independent of the reference band.
\begin{table}[!t]
\caption{Typical co-alignment offsets, in arcsec, (averaged over 7 AIA passbands) measured relative to the waveband listed in the far left column's image positions, for pre-processing under CutOut assumptions only.}
\label{tbl:AIA_PREPROCESS_ALIGNRESULTS_ALLBANDS}
\vspace{0.5mm}
\begin{tabular}{ccccc}
 Band {\AA}  & $\langle X _{CO} \rangle$ & $\sigma ^{X} _{SEM}$ & $\langle Y_{CO} \rangle$ & $\sigma ^{Y} _{SEM}$ \\
 \hline
131  & 0.95	& 1.19	& 1.24	& 1.56 \\
171  & 0.98	& 1.27	& 1.14	& 1.39 \\
193  & 0.96	& 1.21	& 1.19	& 1.47 \\
211  & 1.06	& 1.31	& 1.27	& 1.63 \\
304  & 1.31	& 1.62	& 1.41	& 1.72 \\
335  & 0.97	& 1.22	& 1.56  & 1.89 \\
1600 & 1.73     & 2.32	& 1.08	& 1.32 \\
\hline
\end{tabular}
\end{table}

\section{Discussion}\label{sec:Discussion}

Using an extended three year study, a characterization of AIA (1700\,{\AA}) and HMI's (LOS magnetograms) pointing variations, at sub-weekly time-scales, was performed. Mis-alignments of the two instruments were consistent with previous reports indicating no better than single pixel accuracy exists in the master pointing data \cite{DelZannaetal2011,Brooksetal2012ApJ,Shineetal2011SPD}. Pointing variations of the various AIA passbands were determined to be $\lesssim$\,1.18$\arcsec$ on a daily scale using the recent June 2012 Venus transit, and as such were consistent with the instrument team's expectations ($\lesssim$\,1.2$\arcsec$; \opencite{Shineetal2011SPD})

Two notable artifacts were identified in the long-term pointing data (Figure~\ref{fig:AIA1700_HMI_XYMPS_RES_VSTIME}). The first was a significant shift in AIA's pointing between December 2011 and January 2012, while the second was annual sinusoidal fluctuations in both HMI and AIA's master pointings (Figure~\ref{fig:AIA1700_HMI_XYMPS_RES_VSTIME}). The final artifact was daily variations found in the master pointings of AIA passbands (Figures~\ref{fig:VT_XRESD_TELE_124}, \ref{fig:VT_YRESD_TELE_124}, and \ref{fig:VT_XYRESD_TELE_3}). Below we provide discussions on the causes of these two artifacts as well as how they relate to the information obtained in this article.

The AIA 1700\,{\AA} pointing shift (Figure~\ref{fig:AIA1700_HMI_XYMPS_RES_VSTIME}) is due to adjustments made to telescope 3's thermal control parameters to increase the instrument stability (Boerner 2013, private communication). The heater adjustment occurred on 18 January 2012 and was also performed on telescopes 1 and 2. These types of internal updates have been performed at various times in the mission, and in each case updates are made to master pointings to compensate for the resultant shifts.

The seasonal drift in telescope pointings is likely caused by thermal flexing of the instrument. This flexing is considered to occur between the science telescope ({\it i.e.}, makes passband images), and the guide telescope which drives the image stabilization system (Boerner 2013, private communication). Changes in energy flux due to variations of SDO's distance to the Sun during its yearly orbit, with peak changes correlating with Earth's aphelion and perihelion, are suggested as the origin of the flexing itself. Furthermore, SDO's geosynchronous orbit likely generates these same effects on a daily basis, thereby causing similar daily variations in the master pointings. Temperature variations are believed to be the cause of daily pointing variations ($\approx$\,0.6$\arcsec$\,--\,1.2$\arcsec$; \opencite{Shineetal2011SPD}), to which our results are consistent (Table~\ref{tbl:AIA_PREPROCESS_ALIGNRESULTS_ALLBANDS}).

AIA and HMI long-term pointing variations in the solar $x$ direction, $\approx$\,0.26$\arcsec$, indicates the master pointing accuracy is jitter limited at time-scales below their characterization. This conclusion is further supported by the distribution ({\it i.e.}, random scatter) of our measured $x$ pointings to those of the instrument teams (Figure~\ref{fig:AIA1700_HMI_XYMPS_RES_VSTIME}). It is hypothesized that the consistent nature of the $y$ offsets, between limb and correlation derived pointings, may be an indication of expected satellite flexing or even a payload shift. Finally, based on minimal variations of correlated offsets between AIA and HMI we hypothesize the thermal jitter affecting HMI is similar in scale to AIA's.

\section{Conclusions}\label{sec:concl_summary}

Our study has provided the first characterization of internal pointing errors occurring across SDO's AIA and HMI instruments using cross-correlation techniques at time scales below the team's master pointing data. Long-term correlation offsets provided an estimate of pointing jitter affecting the AIA and HMI instruments. Our solar $x$ pointing variations ($\approx$\,0.3$\arcsec$), between limb and correlation techniques, are within the limits of thermal jitter in the AIA instrument \cite{Lemenetal2012,Aschwandenetal2011}, while those of the solar $y$ direction ($\approx$\,0.5$\arcsec$) are consistent with expected variations \cite{Shineetal2011SPD}. Thermal jitter, as previously highlighted, occurs from thermal bending between the science and guide telescopes as a result of orbit induced temperature variations that occur on yearly and daily time scales. Minimal variations between limb and correlation derived pointings indicate that master pointing data achieve single pixel accuracy on sub-weekly time-scales. Long-term instrument pointing variations between HMI and AIA also suggest that thermal jitter affecting HMI is similar in scale to that of AIA.

From the June 2012 Venus transit, daily pointing variations in all AIA passbands were presented. Our results are consistent with the instrument team's expectations of daily pointing shifts of 1\,--\,2 pixels \cite{Shineetal2011SPD}. The typical co-alignment offsets were $\langle \Delta X_{CO} \rangle =  1.10\arcsec$ and $\langle \Delta Y_{CO} \rangle =  1.25\arcsec$, when aligned to HMI's nominal image center. Finally, these co-alignment accuracies are consistent with previously reported mis-alignments of $\approx$\,1$\arcsec$ in AIA passbands \cite{DelZannaetal2011,Brooksetal2012ApJ}, and expected variations in master pointing data \cite{Shineetal2011SPD}.

\section{Acknowledgements}
The authors are very grateful for discussions and comments from P. Boerner and to R. Shine for providing us with extensive AIA pointing data, both from the AIA team. The authors are very grateful for the referee's constructive comments and suggestions on this article. This research was supported by the National Aeronautics and Space Administration (NASA) grant NNX-07AT01G. N. B. Orange was also supported by the Florida Space Grant Consortium, a NASA sponsored program administered by the University of Central Florida, grant NNX-10AM01H. Any opinions, findings, and conclusions or recommendations expressed in this material are those of the author(s) and do not necessarily reflect the views of the NSF or NASA. The final publication is available at http://link.springer.com/article/10.1007\%2Fs11207-013-0441-2.

\bibliographystyle{spr-mp-sola}
\bibliography{vtstudy}
\end{article}
\end{document}